\documentclass[global]{svjour}
\usepackage{amssymb}
\usepackage{amsfonts}
\usepackage{amsmath}

\input{tcilatex}

\begin{document}

\journalname{In print on \emph{Int. J. Non-Linear Mechanics}}
\title{$SO\left( 3\right) $ \textbf{INVARIANCE\ AND COVARIANCE IN\ MIXTURES
OF SIMPLE BODIES}}
\author{\textbf{Paolo Maria Mariano}}
\institute{Dipartimento di Ingegneria Strutturale e Geotecnica,\\
Universit\`{a} di Roma "La Sapienza,\\
via Eudossiana 18, 00184 Roma (Italy),\\
%TCIMACRO{\TeXButton{email}{\email{paolo.mariano@uniroma1.it}}}%
%BeginExpansion
\email{paolo.mariano@uniroma1.it}%
%EndExpansion
\\
and Universit\`{a} del Molise, Campobasso (Italy).}
\dedication{}
\maketitle

\begin{abstract}
We adapt to mixtures the procedure of invariance of external power under the
action of $SO\left( 3\right) $ to deduce balance equations. The two
classical axioms about the growths of momentum and moment of momentum are
derived with the help of a rule on the structure of the total power. We
discuss also the covariance of the balance equations of each constituent and
the nature of the constituent stress.
\end{abstract}


\section{Commentary to the theory of mixtures of simple bodies: the r\^{o}le
of the invarance of the power.}

\subsection{Preliminary remarks.}

Let $\mathcal{B}_{0}$\ be the regular (in the sense of "fit") region of the
three-dimensional Euclidean point space $\mathcal{E}^{3}$ occupied by the
mixture in a reference place. We call \emph{part} of $\mathcal{B}_{0}$\ any
regular subset $\mathfrak{b}$ of it and denote with $\mathbf{X}$ its generic
point. The regular region $\mathcal{B}$\ occupied by the mixture in the
current place is obtained from $\mathcal{B}_{0}$\ through a sufficiently
smooth orientation preserving mapping $f$ such that $f\left( \mathcal{B}%
_{0}\right) =\mathcal{B}$; for any $\mathbf{x\in }\mathcal{B}$ we have then $%
\mathbf{x=}f\left( \mathbf{X}\right) $. Motions are sufficiently smooth
one-parameter mappings $f_{t}$\textbf{,} with $t\in \left[ t_{0},t_{1}\right]
$ the time. The velocity in the spatial description is denoted with $\mathbf{%
v}$, while the acceleration with $\mathbf{a}=\partial _{t}\mathbf{v}+\left(
grad\mathbf{v}\right) \mathbf{v}$.

At the introduction of Lecture 5 of his \textquotedblleft Rational
Thermodynamics\textquotedblright , Truesdell [1] describes the mixtures by
assuming as a crucial point of view that all constituents occupy $\mathcal{B}
$ \emph{simultaneously}. Such an assumption is not true at a `small' scale%
\footnote{%
The physical meaning of the adjective `small' changes for each type of
mixture.} but it can be considered valid at the coarse scale that allows us
to use the standard procedures of continuum mechanics. In this way, each
point $\mathbf{x}$ of $\mathcal{B}$ has indeed the properties of all the
constituents: with $\mathbf{x}_{\alpha }^{\prime }\left( \mathbf{x,}t\right) 
$ we indicate the \emph{velocity }of the $\alpha $th constituent at $\mathbf{%
x}$ (a spatial field), while with $\mathbf{x}_{\alpha }^{\prime \prime }$
the acceleration of the same constituent at $\mathbf{x}$ (the apex means
here time derivative); the \emph{density of mass} of the same constituent is 
$\rho _{\alpha }$, and $c_{\alpha }$ the concentration of mass, i.e. the
ratio $c_{\alpha }=\frac{\rho _{\alpha }}{\rho }$, where $\rho =\sum_{\alpha
}\rho _{\alpha }$. When we consider the mixture as a single body, the
velocity $\mathbf{v}$ at $\mathbf{x}$ and $t$ is given by the sum of the
velocities of each constituent weighed by the concentrations: 
\begin{equation}
\mathbf{v=}\sum_{\alpha }c_{\alpha }\mathbf{x}_{\alpha }^{\prime }.
\label{1}
\end{equation}%
Note that when all the constituents have the same velocity $\mathbf{x}%
^{\prime }=\mathbf{x}_{\alpha }^{\prime }$ for any $\alpha $, then $\mathbf{%
v=x}^{\prime }$ because $\sum_{\alpha }c_{\alpha }=1$.

An external spatial observer calls \emph{rigid} all the velocity fields $%
\mathbf{v}_{R}$ (the subscript R stands for `rigid') of the form 
\begin{equation}
\mathbf{v}_{R}=\mathbf{c}\left( t\right) +\mathbf{\dot{q}\times }\left( 
\mathbf{x}-\mathbf{x}_{0}\right) ,  \label{2}
\end{equation}%
where $\mathbf{c}\left( t\right) $\ and $\mathbf{\dot{q}}\left( t\right) $\
are translational and\ rotational velocities respectively. At a certain
instant $t$, a spatial observer evaluates a velocity $\mathbf{v}$ (and $%
\mathbf{x}_{\alpha }^{\prime }$ for the $\alpha $th constituent). If the
observer changes its own state, new velocities $\mathbf{v}^{\ast }$\ and $%
\mathbf{x}_{\alpha }^{\prime \ast }$ are measured and the values $\mathbf{v}$
and $\mathbf{v}^{\ast }$\ (or $\mathbf{x}_{\alpha }^{\prime }$\ and $\mathbf{%
x}_{\alpha }^{\prime \ast }$) are related by the rules describing the change
of state of the observer. We consider first changes of observer ruled by
elements of the special orthogonal group $SO\left( 3\right) $ (the observers
differ trough a rigid body motion). We then have%
\begin{equation}
\mathbf{x}_{\alpha }^{\prime \ast }=\mathbf{x}_{\alpha }^{\prime }\mathbf{+c}%
\left( t\right) +\mathbf{\dot{q}\times }\left( \mathbf{x}-\mathbf{x}%
_{0}\right) ,  \label{3}
\end{equation}%
which includes both Galilean ($\mathbf{c}$ constant in time and $\mathbf{%
\dot{q}=0}$) and rotational ($\mathbf{c=0}$, $\mathbf{\ddot{q}=0}$) changes
of observer.

Equation (\ref{3}) implies that \emph{each observer sees all the constituents%
}. This justify the choice of common translational and rotational
velocities. Each observer is, in fact, a slicing of space-time. The
simultaneous presence of constituents can be viewed as a simultaneous
presence of slicings. With (\ref{3}) we then presume that all slicings are
isometrically `deformed' with the same translational and rotational
velocities.

\subsection{The main result.}

The picture of the simultaneous presence of all the constituents at each
point $\mathbf{x}$ of $\mathcal{B}$ is enriched naturally when we try to
represent the interactions within the mixture. The additional ingredient
with respect to the mechanics of simple bodies is, in fact, the need to
account for the interactions between each constituent and the others and to
represent them in some way. Truesdell suggested three methaphysical
principles for mixtures (see [1]): (i) the properties of the whole mixture
are consequences of the ones of the constituents, (ii) each of them can be
isolated from the rest provided that the interactions with the others are
accounted for, (iii) the whole mixture behaves as a single body.

When these principles are applied to evaluate the \emph{balance of mass},
taking into account (ii) we have 
\begin{equation}
\rho \breve{c}_{\alpha }=\partial _{t}\rho _{\alpha }+div\left( \rho
_{\alpha }\mathbf{x}_{\alpha }^{\prime }\right)  \label{4}
\end{equation}%
for the $\alpha $th constituent, where $\breve{c}_{\alpha }$\ is the \emph{%
growth of mass} of the same constituent. On the other hand, if the mixture
is isolated, from the principles (i), (ii) and (iii) (chiefly from principle
(iii)) we obtain by summation of (\ref{4}) on $\alpha $%
\begin{equation}
\dot{\rho}+\rho div\mathbf{v}=0\text{ \ \ \ \ \ \ in }\mathcal{B},  \label{5}
\end{equation}%
\begin{equation}
\sum_{\alpha }\breve{c}_{\alpha }=0.  \label{6}
\end{equation}

To deduce the balance equations for a mixture, we adapt here the procedure
based on a request of invariance of the external power under changes of
observers developed in [2]. This allows us to construct a tool useful for
deriving models of mixtures of complex bodies. The basic outline of such a
procedure is as follows:

\begin{itemize}
\item We write first the power developed by all external interactions on an
arbitrary part of a certain constituent and include explicitly terms of
interaction with the other constituents because, though all costituents are
in the same place, each of them is "external" with respect to the others. A
requirement of invariance of the power with respect to changes of observer
ruled by (\ref{3}) allows us to obtain integral (and - by localization -
pointwise) balance equations of momentum and moment of momentum.

\item We write also the external power on the same part of the mixture,\
considered globally in the spirit of principle (iii) and \emph{assume} that
it is the sum of the powers of the constituents calculated on the velocity
of the whole mixture, for any choice of $\mathbf{v}$ (and we call this
assumption \textquotedblleft \emph{rule of the total power}%
\textquotedblright\ for convenience).
\end{itemize}

The main result of this procedure is the proof that the growth of momentum
on a given constituent due to the others is self-equilibrated, and the same
holds for the growth of moment of momentum, while in the standard
presentation of the theory these properties are assumed as axioms.

Let $\mathfrak{b}$ be any arbitrary part of the body, we write the power $%
\mathcal{P}_{\mathfrak{b}}^{\alpha }$ developed on the $\alpha $th\
constituent occupying $\mathfrak{b}$\ as 
\begin{equation}
\mathcal{P}_{\mathfrak{b}}^{\alpha }=\int_{\mathfrak{b}}\rho _{\alpha }%
\mathbf{b}_{\alpha }\cdot \mathbf{x}_{\alpha }^{\prime }+\int_{\partial 
\mathfrak{b}}\mathbf{T}_{\alpha }\mathbf{n}\cdot \mathbf{x}_{\alpha
}^{\prime }+\int_{\mathfrak{b}}\rho \mathbf{\check{m}}_{\alpha }\cdot 
\mathbf{x}_{\alpha }^{\prime }+\int_{\mathfrak{b}}\rho \mathbf{\check{\mu}}%
_{\alpha }\cdot curl\mathbf{x}_{\alpha }^{\prime },  \label{7}
\end{equation}%
where $\mathbf{n}$\ is the outward unit normal to the boundary $\partial 
\mathfrak{b}$\ of $\mathfrak{b}$, $\mathbf{b}_{\alpha }$\ the vector of body
forces acting on the $\alpha $th\ constituent, $\mathbf{T}_{\alpha }$\ the
Cauchy stress tensor on the same constituent, $\mathbf{\check{m}}_{\alpha }$%
\ a vector representing the \emph{growth of momentum} induced by the other
constituents on the $\alpha $th one, $\mathbf{\check{\mu}}_{\alpha }$\ the
analogous \emph{growth of moment of momentum}. In the spirit of principle
(ii), the last two integrals in (\ref{7}) represent the power of the direct
interactions between the $\alpha $th\ constituent and the others.

The principle (iii) suggests also to write the power on a part $\mathfrak{b}$
occupied by the whole mixture as the power of a simple body. As a
consequence, we write the standard relation 
\begin{equation}
\mathcal{P}_{\mathfrak{b}}=\int_{\mathfrak{b}}\rho \mathbf{b}\cdot \mathbf{v}%
+\int_{\partial \mathfrak{b}}\mathbf{Tn}\cdot \mathbf{v,}  \label{8}
\end{equation}%
where $\mathbf{b}$ is the vector of the total bulk interactions and $\mathbf{%
T}$ the total Cauchy stress.

At each $\mathbf{x}$, $\mathbf{b}$ and $\mathbf{b}_{\alpha }$\ and $\mathbf{T%
}$ and $\mathbf{T}_{\alpha }$\ are related (see [1]) by 
\begin{equation}
\mathbf{b=}\sum_{\alpha }c_{\alpha }\mathbf{b}_{\alpha },  \label{9}
\end{equation}
\begin{equation}
\mathbf{T=}\sum_{\alpha }\mathbf{T}_{\alpha }.  \label{10}
\end{equation}%
Note that the bulk measures of interaction $\mathbf{b}$ and $\mathbf{b}%
_{\alpha }$ are comprehensive of inertial actions and we consider valid the
additive decomposition into the inertial ($in$) and non-inertial ($ni$)
components, namely%
\begin{equation}
\mathbf{b}_{\alpha }=\mathbf{b}_{\alpha }^{ni}+\mathbf{b}_{\alpha }^{in},
\label{11}
\end{equation}%
\begin{equation}
\mathbf{b}=\mathbf{b}^{ni}+\mathbf{b}^{in}.  \label{12}
\end{equation}

\begin{theorem}
In the theory of mixtures of simple bodies, the two groups of assertions
listed in what follows are equivalent.
\end{theorem}

\emph{First group}.

\begin{itemize}
\item $\mathcal{P}_{\mathfrak{b}}^{\alpha }$ \emph{is invariant under
classical changes of observer ruled by (\ref{3}), i.e.} 
\begin{equation}
\mathcal{P}_{\mathfrak{b}}^{\alpha }\left( \mathbf{x}_{\alpha }^{\prime \ast
}\right) =\mathcal{P}_{\mathfrak{b}}^{\alpha }\left( \mathbf{x}_{\alpha
}^{\prime }\right)  \label{13}
\end{equation}%
\emph{for any choice of }$\mathfrak{b}$, $\mathbf{c}\left( t\right) $ \emph{%
and} $\mathbf{\dot{q}}\left( t\right) $.

\item \emph{The following rule of the kinetic energy and the power of
inertial forces holds:} 
\begin{equation*}
\frac{d}{dt}\int_{\mathfrak{b}}\frac{1}{2}\rho _{\alpha }\mathbf{x}_{\alpha
}^{\prime }\cdot \mathbf{x}_{\alpha }^{\prime }-\frac{1}{2}\int_{\partial 
\mathfrak{b}}\rho _{\alpha }\mathbf{x}_{\alpha }^{\prime }\cdot \mathbf{x}%
_{\alpha }^{\prime }\left( \mathbf{x}_{\alpha }^{\prime }\cdot \mathbf{n}%
\right) +
\end{equation*}%
\begin{equation}
+\frac{1}{2}\int_{\mathfrak{b}}\rho \breve{c}_{\alpha }\mathbf{x}_{\alpha
}^{\prime }\cdot \mathbf{x}_{\alpha }^{\prime }+\int_{\mathfrak{b}}\rho
_{\alpha }\mathbf{b}_{\alpha }^{in}\cdot \mathbf{x}_{\alpha }^{\prime }=0
\label{14}
\end{equation}%
\emph{for any} $\mathfrak{b}$.

\item \emph{The rule of the total power holds:} 
\begin{equation}
\mathcal{P}_{\mathfrak{b}}\left( \mathbf{v}\right) =\sum_{\alpha }\mathcal{P}%
_{\mathfrak{b}}^{\alpha }\left( \mathbf{v}\right)  \label{15}
\end{equation}%
\emph{for any choice of }$\mathbf{v}$\emph{.}
\end{itemize}

\emph{Second group.}

\begin{itemize}
\item \emph{For each constituent, the following equations of balance of
momentum and moment of momentum hold:} 
\begin{equation}
\rho _{\alpha }\mathbf{b}_{\alpha }^{ni}+div\mathbf{T}_{\alpha }+\rho 
\mathbf{\check{m}}_{\alpha }=\rho c_{\alpha }\mathbf{x}_{\alpha }^{\prime
\prime }+\rho \breve{c}_{\alpha }\mathbf{x}_{\alpha }^{\prime },  \label{16}
\end{equation}%
\begin{equation}
\mathbf{T}_{\alpha }-\mathbf{T}_{\alpha }^{T}=-\rho \mathbf{\check{\mu}}%
_{\alpha }\times .  \label{17}
\end{equation}

\item \emph{The growths of momentum due to the interactions between
constituents are pointwise self-equilibrated:} 
\begin{equation}
\sum_{\alpha }\mathbf{\check{m}}_{\alpha }=0.  \label{18}
\end{equation}

\item \emph{The growths of moment of momentum due to the interactions
between constituents are pointwise self-equilibrated:} 
\begin{equation}
\sum_{\alpha }\mathbf{\check{\mu}}_{\alpha }=0.  \label{19}
\end{equation}

\item \emph{For the whole mixture, the following equations of balance of
momentum and moment of momentum hold:} 
\begin{equation}
\rho \mathbf{b}^{ni}+div\left( \mathbf{T+\sum_{\alpha }}c\mathbf{_{\alpha }u}%
_{\alpha }\otimes \mathbf{u}_{\alpha }\right) =\rho \mathbf{a,}  \label{20}
\end{equation}%
\begin{equation}
skw\mathbf{T=0,}  \label{21}
\end{equation}%
\emph{where} $\mathbf{u}_{\alpha }=\mathbf{x}_{\alpha }^{\prime }-\mathbf{v}$
\emph{is the velocity of diffusion of the }$\alpha $\emph{th\ constituent}.
\end{itemize}

In (\ref{20}), the term $\mathbf{\sum_{\alpha }}c\mathbf{_{\alpha }u}%
_{\alpha }\otimes \mathbf{u}_{\alpha }$ accounts for the diffusive motion of
the constituents of the mixture. Analogous terms are present in the kinetic
theory of gases.

\subsection{Proof of Theorem 1.}

We prove here only that the second group of assertions follows from the
first group, because the converse is rather immediate.

First we consider the power $\mathcal{P}_{\mathfrak{b}}^{\alpha }$\ and
realize that the request of invariance (13) is satisfied if and only if%
\begin{equation*}
\mathbf{c}\cdot \left( \int_{\mathfrak{b}}\rho _{\alpha }\mathbf{b}_{\alpha
}+\int_{\partial \mathfrak{b}}\mathbf{T}_{\alpha }\mathbf{n}+\int_{\mathfrak{%
b}}\rho \mathbf{\check{m}}_{\alpha }\right) +
\end{equation*}%
\begin{equation*}
+\mathbf{\dot{q}\cdot }\left( \int_{\mathfrak{b}}\left( \mathbf{x}-\mathbf{x}%
_{0}\right) \mathbf{\times }\rho _{\alpha }\mathbf{b}_{\alpha
}+\int_{\partial \mathfrak{b}}\left( \mathbf{x}-\mathbf{x}_{0}\right) 
\mathbf{\times T}_{\alpha }\mathbf{n}\right) +
\end{equation*}%
\begin{equation}
+\mathbf{\dot{q}\cdot }\left( \int_{\mathfrak{b}}\left( \mathbf{x}-\mathbf{x}%
_{0}\right) \mathbf{\times }\rho \mathbf{\check{m}}_{\alpha }+\int_{%
\mathfrak{b}}\rho \mathbf{\check{\mu}}_{\alpha }\right) =0,  \label{22}
\end{equation}%
for any choice of $\mathbf{c}$\ and $\mathbf{\dot{q}}$, whose arbitraryness
implies the integral balances of forces and couples for the $\alpha $th
constituent:%
\begin{equation}
\int_{\mathfrak{b}}\rho _{\alpha }\mathbf{b}_{\alpha }+\int_{\partial 
\mathfrak{b}}\mathbf{T}_{\alpha }\mathbf{n}+\int_{\mathfrak{b}}\rho \mathbf{%
\check{m}}_{\alpha }=0,  \label{23}
\end{equation}%
\begin{equation}
\int_{\mathfrak{b}}\left( \mathbf{x}-\mathbf{x}_{0}\right) \mathbf{\times }%
\rho _{\alpha }\mathbf{b}_{\alpha }+\int_{\partial \mathfrak{b}}\left( 
\mathbf{x}-\mathbf{x}_{0}\right) \mathbf{\times T}_{\alpha }\mathbf{n}+\int_{%
\mathfrak{b}}\left( \mathbf{x}-\mathbf{x}_{0}\right) \mathbf{\times }\rho 
\mathbf{\check{m}}_{\alpha }+\int_{\mathfrak{b}}\rho \mathbf{\check{\mu}}%
_{\alpha }=0.  \label{24}
\end{equation}%
The arbitraryness of $\mathfrak{b}$ and the use of the Gauss theorem allow
us to obtain from (23) the pointwise balance%
\begin{equation}
\rho _{\alpha }\mathbf{b}_{\alpha }+div\mathbf{T}_{\alpha }+\rho \mathbf{%
\check{m}}_{\alpha }=0,  \label{25}
\end{equation}%
while equation (\ref{17}) follows from (\ref{24}) with the additional use of
(\ref{25}).

A standard transport theorem and the use of the balance of mass (\ref{4})
allow us to calculate the time derivative of the first integral in (\ref{14}%
): we then have%
\begin{equation}
\frac{1}{2}\frac{d}{dt}\int_{\mathfrak{b}}\rho _{\alpha }\mathbf{x}_{\alpha
}^{\prime }\cdot \mathbf{x}_{\alpha }^{\prime }=\int_{\mathfrak{b}}\rho
_{\alpha }\mathbf{x}_{\alpha }^{\prime \prime }\cdot \mathbf{x}_{\alpha
}^{\prime }+\frac{1}{2}\int_{\partial \mathfrak{b}}\frac{1}{2}\rho _{\alpha }%
\mathbf{x}_{\alpha }^{\prime }\cdot \mathbf{x}_{\alpha }^{\prime }\left( 
\mathbf{x}_{\alpha }^{\prime }\cdot \mathbf{n}\right) +\int_{\mathfrak{b}}%
\frac{1}{2}\rho \breve{c}_{\alpha }\mathbf{x}_{\alpha }^{\prime }\cdot 
\mathbf{x}_{\alpha }^{\prime }  \label{26}
\end{equation}%
By inserting (\ref{26}) in (\ref{14}) we obtain%
\begin{equation}
\int_{\mathfrak{b}}\left( \rho _{\alpha }\mathbf{x}_{\alpha }^{\prime \prime
}+\rho \breve{c}_{\alpha }\mathbf{x}_{\alpha }^{\prime }+\rho _{\alpha }%
\mathbf{b}_{\alpha }^{in}\right) \cdot \mathbf{x}_{\alpha }^{\prime }=0.
\label{27}
\end{equation}%
Even in this case both $\mathfrak{b}$ and $\mathbf{x}_{\alpha }^{\prime }$\
are arbitrary, then%
\begin{equation}
\rho _{\alpha }\mathbf{b}_{\alpha }^{in}=-\rho _{\alpha }\mathbf{x}_{\alpha
}^{\prime \prime }-\rho \breve{c}_{\alpha }\mathbf{x}_{\alpha }^{\prime }.
\label{28}
\end{equation}%
By substituting (\ref{28}) and (\ref{11}) in (\ref{25}), the pointwise
balance of momentum of the $\alpha $th constituent (\ref{16}) follows.

Since the rule of the total power holds for any choice of $\mathbf{v}$, it
holds for rigid velocity fields of the form (\ref{2}). As a consequence,
from (\ref{15}) we obtain%
\begin{equation*}
\mathbf{c}\cdot \left( \int_{\mathfrak{b}}\rho \mathbf{b}+\int_{\partial 
\mathfrak{b}}\mathbf{Tn}\right) +\mathbf{\dot{q}\cdot }\left( \int_{%
\mathfrak{b}}\left( \mathbf{x}-\mathbf{x}_{0}\right) \mathbf{\times }\rho 
\mathbf{b}+\int_{\partial \mathfrak{b}}\left( \mathbf{x}-\mathbf{x}%
_{0}\right) \mathbf{\times Tn}\right) =
\end{equation*}%
\begin{equation*}
=\mathbf{c}\cdot \left( \int_{\mathfrak{b}}\sum_{\alpha }\left( \rho
_{\alpha }\mathbf{b}_{\alpha }+\rho \mathbf{\check{m}}_{\alpha }\right)
+\int_{\partial \mathfrak{b}}\sum_{\alpha }\mathbf{T}_{\alpha }\mathbf{n}%
\right) +
\end{equation*}%
\begin{equation*}
+\mathbf{\dot{q}\cdot }\left( \int_{\mathfrak{b}}\sum_{\alpha }\left( 
\mathbf{x}-\mathbf{x}_{0}\right) \mathbf{\times }\left( \rho _{\alpha }%
\mathbf{b}_{\alpha }+\rho \mathbf{\check{m}}_{\alpha }\right) \right) +
\end{equation*}%
\begin{equation}
+\mathbf{\dot{q}\cdot }\left( \int_{\partial \mathfrak{b}}\sum_{\alpha
}\left( \mathbf{x}-\mathbf{x}_{0}\right) \mathbf{\times T}_{\alpha }\mathbf{%
n+}\int_{\mathfrak{b}}\sum_{\alpha }\rho \mathbf{\check{\mu}}_{\alpha
}\right) ,  \label{29}
\end{equation}%
i.e.%
\begin{equation}
\int_{\mathfrak{b}}\rho \mathbf{b}+\int_{\partial \mathfrak{b}}\mathbf{Tn=}%
\int_{\mathfrak{b}}\sum_{\alpha }\left( \rho _{\alpha }\mathbf{b}_{\alpha
}+\rho \mathbf{\check{m}}_{\alpha }\right) +\int_{\partial \mathfrak{b}%
}\sum_{\alpha }\mathbf{T}_{\alpha }\mathbf{n,}  \label{30}
\end{equation}%
\begin{equation*}
\int_{\mathfrak{b}}\left( \mathbf{x}-\mathbf{x}_{0}\right) \mathbf{\times }%
\rho \mathbf{b}+\int_{\partial \mathfrak{b}}\left( \mathbf{x}-\mathbf{x}%
_{0}\right) \mathbf{\times Tn=}
\end{equation*}%
\begin{equation}
=\int_{\mathfrak{b}}\sum_{\alpha }\left( \mathbf{x}-\mathbf{x}_{0}\right) 
\mathbf{\times }\left( \rho _{\alpha }\mathbf{b}_{\alpha }+\rho \mathbf{%
\check{m}}_{\alpha }\right) +\int_{\partial \mathfrak{b}}\sum_{\alpha
}\left( \mathbf{x}-\mathbf{x}_{0}\right) \mathbf{\times T}_{\alpha }\mathbf{%
n+}\int_{\mathfrak{b}}\sum_{\alpha }\rho \mathbf{\check{\mu}}_{\alpha }.
\label{31}
\end{equation}%
By inserting (\ref{9}) and (\ref{10}) in (\ref{30}) and (\ref{31}), we
obtain (\ref{18}) and (\ref{19}) thanks to the arbitraryness of $\mathfrak{b}
$.

The balance equations (\ref{20}) and (\ref{21}) of the whole mixture follow
by summing over $\alpha $\ equations (\ref{16}) and (\ref{17}); we make use
of (\ref{18}) and (\ref{19}) and take into account that for any spatial
field $\lambda \left( \mathbf{x,}t\right) =\sum_{\alpha }$ $c_{\alpha
}\lambda _{\alpha }\left( \mathbf{x,}t\right) $ we have (see [1])%
\begin{equation}
\dot{\lambda}=\sum_{\alpha }c_{\alpha }\lambda _{\alpha }^{\prime }-\rho
^{-1}div\sum_{\alpha }c_{\alpha }\lambda _{\alpha }\mathbf{u}_{\alpha
}+\sum_{\alpha }\breve{c}_{\alpha }\lambda _{\alpha }.  \label{32}
\end{equation}%
$\square $

\textbf{Remark} (\emph{about boundary conditions}). The problem of assigning
boundary conditions in mixtures is in general rather delicate (and crucial)
because we know the total boundary traction $\mathbf{Tn=t}$ and do not know
the constituent traction $\mathbf{T}_{\alpha }\mathbf{n=t}_{\alpha }$. The
question has been discussed in detail by Rajagopal and co-workers (here we
do not tackle the problem) and the related results (together with
comparisons with the proposals in the existing literature) can be found in
[3] (see also [4, 5]).

\section{A theorem of the kinetic energy and the power of inertial forces
for the whole mixture.}

In the present Section we reverse the point of view followed so far: we do
not consider the single constituent but account for the whole mixture by
connecting (\ref{8}) with (\ref{20}) through the use of an appropriate
theorem of the kinetic energy and the power of inertial forces.

Basically, we consider the term $div\mathbf{\sum_{\alpha }}c\mathbf{_{\alpha
}u}_{\alpha }\otimes \mathbf{u}_{\alpha }$ of inertial nature and associate
it with $\mathbf{b}^{in}$. This interpretation is not standard. The
decomposition (\ref{11}) is in general adopted to separate the objective
part of the bulk interactions (namely $\mathbf{b}^{ni}$) from the rest: the
inertial terms are, in fact, not objective. Here the term $\mathbf{u}%
_{\alpha }\otimes \mathbf{u}_{\alpha }$ is objective, so we consider $%
\mathbf{b}^{in}$ endowed with an objective part.

The analogous expression of (\ref{3}) for $\mathbf{v}$ is%
\begin{equation}
\mathbf{v}^{\ast }=\mathbf{v+c}\left( t\right) +\mathbf{\dot{q}\times }%
\left( \mathbf{x-x}_{0}\right) .  \label{33}
\end{equation}%
The values $\mathbf{v}$ and $\mathbf{v}^{\ast }$ are the ones measured
respectively before and after "rigid" changes of observer ruled by $SO\left(
3\right) $. If we require the invariance of $\mathcal{P}_{\mathfrak{b}}$%
\textbf{\ }in (\ref{8}) under (\ref{33}), we obtain the standard pointwise
balance%
\begin{equation}
\rho \mathbf{b+}div\mathbf{T=0},  \label{34}
\end{equation}%
in which (\ref{11}) applies. Taking into account (10), the compatibility of (%
\ref{34}) with (\ref{20}) is assured if we postulate for the whole mixture
that%
\begin{equation*}
\frac{d}{dt}\int_{\mathfrak{b}}\frac{1}{2}\rho \mathbf{v}\cdot \mathbf{v}%
+\int_{\partial \mathfrak{b}}\sum_{\alpha }c_{\alpha }\mathbf{u}_{\alpha
}\cdot \mathbf{v}\left( \mathbf{u}_{\alpha }\cdot \mathbf{n}\right) +
\end{equation*}%
\begin{equation}
+\int_{\mathfrak{b}}\mathbf{\sum_{\alpha }}c\mathbf{_{\alpha }u}_{\alpha
}\otimes \mathbf{u}_{\alpha }\cdot grad\mathbf{v}+\int_{\mathfrak{b}}\rho 
\mathbf{b}^{in}\cdot \mathbf{v}=0  \label{35}
\end{equation}%
for any \emph{fixed} control volume $\mathfrak{b}$ in $\mathcal{B}$.

\begin{itemize}
\item Since $\mathfrak{b}$ is fixed, we interpret%
\begin{equation}
\int_{\partial \mathfrak{b}}\sum_{\alpha }c_{\alpha }\mathbf{u}_{\alpha
}\cdot \mathbf{v}\left( \mathbf{u}_{\alpha }\cdot \mathbf{n}\right)
\label{36}
\end{equation}%
as the total flow of kinetic energy through the boundary $\partial \mathfrak{%
b}$ due to the relative motion of the constituents with respect to $%
\mathfrak{b}$.

\item Moreover, the term%
\begin{equation}
\int_{\mathfrak{b}}\mathbf{\sum_{\alpha }}c\mathbf{_{\alpha }u}_{\alpha
}\otimes \mathbf{u}_{\alpha }\cdot grad\mathbf{v}  \label{37}
\end{equation}%
is interpreted as a \emph{production} of kinetic energy due to the
cooperation of the non-uniformity of the velocity of the whole mixture and
the local relative agitation of the constituents.
\end{itemize}

\section{The first and the second principle of thermodynamics}

For the first principle in $\mathfrak{b}$, we simply add a "source" term to
the standard expression. This term is due to the simultaneous presence of
all constituents in $\mathfrak{b}$, which is assumed to be fixed. By
indicating with $\varepsilon _{\alpha }$ the density of internal energy of
the $\alpha $th constituent and with $E_{\mathfrak{b}}^{\alpha }$ the supply
of energy in $\mathfrak{b}$ due to the other constituents, we write%
\begin{equation}
\frac{d}{dt}\int_{\mathfrak{b}}\rho _{\alpha }\varepsilon _{\alpha }-E_{%
\mathfrak{b}}^{\alpha }-\mathcal{P}_{\mathfrak{b}}^{\alpha }\left( \mathbf{x}%
_{\alpha }^{\prime }\right) -\int_{\partial \mathfrak{b}}\mathbf{q}_{\alpha
}\cdot \mathbf{n}-\int_{\mathfrak{b}}\rho _{\alpha }r_{\alpha }=0,
\label{38}
\end{equation}%
where $\mathbf{q}_{\alpha }$\ is the density of heat flux through the $%
\alpha $th constituent and $r_{\alpha }$\ the relevant heat source. For $E_{%
\mathfrak{b}}^{\alpha }$, the simplest choice is%
\begin{equation}
E_{\mathfrak{b}}^{\alpha }=\int_{\mathfrak{b}}\rho \breve{\varepsilon}%
_{\alpha },  \label{38bis}
\end{equation}%
where $\breve{\varepsilon}_{\alpha }$\ is the \emph{growth of internal energy%
} induced on the $\alpha $th constituent by the other ones. By inserting (%
\ref{7}) and (\ref{38bis}) in (\ref{38}) and taking into account the
arbitraryness of $\mathfrak{b}$, we obtain the pointwise balance of energy%
\begin{equation}
\rho _{\alpha }\varepsilon _{\alpha }^{\prime }+\rho \breve{c}_{\alpha
}\varepsilon _{\alpha }-\rho \breve{\varepsilon}_{\alpha }-\mathbf{T}%
_{\alpha }\cdot symgrad\mathbf{x}_{\alpha }^{\prime }-div\mathbf{q}_{\alpha
}-\rho _{\alpha }r_{\alpha }=0,  \label{38ter}
\end{equation}%
which has been suggested in [6] ($sym$ extracts the symmetric part of its
argument).

To render (\ref{38}) compatible with the pointwise balance of energy
suggested by Truesdell (see [1] and [3]), we should assume for $E_{\mathfrak{%
b}}^{\alpha }$ the following expression:%
\begin{equation}
E_{\mathfrak{b}}^{\alpha }=\int_{\mathfrak{b}}\left( \rho \breve{\varepsilon}%
_{\alpha }+\frac{1}{2}\rho \breve{c}_{\alpha }\mathbf{x}_{\alpha }^{\prime
}\cdot \mathbf{x}_{\alpha }^{\prime }-\rho \mathbf{\check{m}}_{\alpha }\cdot 
\mathbf{x}_{\alpha }^{\prime }-\rho \mathbf{\check{\mu}}_{\alpha }\cdot curl%
\mathbf{x}_{\alpha }^{\prime }\right) .  \label{39}
\end{equation}%
In writing (\ref{39}) we would presume that, given a certain constituent $%
\alpha $ in $\mathfrak{b}$, the others there have a twofold nature: from one
hand they are external to the $\alpha $th\ one (from which the presence of
the interaction terms in (\ref{7})), while from the other hand, since all
the constituents \emph{occupy} $\mathfrak{b}$, the interaction terms in\ (%
\ref{7}) contribute also to the supply of the internal energy. In other
words, the supply of power furnished by the other constituents would be also
interpreted as a supply of internal energy. In this way, by using (\ref{7}),
(\ref{39}) and developing the time derivative in (\ref{38}), the
arbitraryness of $\mathfrak{b}$ implies the pointwise balance%
\begin{equation*}
\rho _{\alpha }\varepsilon _{\alpha }^{\prime }+\rho \breve{c}_{\alpha
}\varepsilon _{\alpha }-\rho \breve{\varepsilon}_{\alpha }+\rho \mathbf{%
\check{m}}_{\alpha }\cdot \mathbf{x}_{\alpha }^{\prime }-\mathbf{T}_{\alpha
}\cdot grad\mathbf{x}_{\alpha }^{\prime }-\frac{1}{2}\rho \breve{c}_{\alpha }%
\mathbf{x}_{\alpha }^{\prime }\cdot \mathbf{x}_{\alpha }^{\prime }-
\end{equation*}%
\begin{equation}
-div\mathbf{q}_{\alpha }-\rho _{\alpha }r_{\alpha }=0,  \label{40}
\end{equation}%
which is the one in [1] and [3].

The first principle for the whole mixture is deduced by summation over $%
\alpha $, as a consequence of principle (iii). In developing the summation,
following [1], one assumes as axiom that%
\begin{equation}
\mathbf{\sum_{\alpha }}\breve{\varepsilon}_{\alpha }=0.  \label{41}
\end{equation}

For the second principle, we have the standard expression%
\begin{equation}
\frac{d}{dt}\int_{\mathfrak{b}}\rho _{\alpha }\eta _{\alpha }\geq \int_{%
\mathfrak{b}}\rho \breve{\eta}_{\alpha }+\int_{\partial \mathfrak{b}}\mathbf{%
h}_{\alpha }-\int_{\mathfrak{b}}\rho _{\alpha }s_{\alpha },  \label{42}
\end{equation}%
where $\eta _{\alpha }$\ is the density of entropy in the $\alpha $th
constituent, $\breve{\eta}_{\alpha }$\ the growth of entropy induced by the
other constituents, $\mathbf{h}_{\alpha }$\ and $s_{\alpha }$\ entropy flux
and source respectively. The pointwise thermodynamic relations derived and
the integral ones may be exploited by making use of basic results in
hermodynamics of mixtures [3, 7-9].

\section{Covariance}

Each observer is a special representation of space and time. Covariance
means invariance with respect to changes of observer ruled by sufficiently
smooth time-parametrized families of diffeomorphisms $\zeta _{t}:\mathcal{E}%
^{3}\rightarrow \mathcal{E}^{3}$ such that $\zeta _{0}=id$ ($id$ means
identity).

In particular, we consider $n$ syncronous copies of the same slicing ($n$ is
the number of constituents); each copy is occupied by one constituent. To
analize the covariance of the balance equations of the $\alpha $th
constituent, we consider an arbitrary $\zeta _{t}^{\alpha }$\ acting only on
the slicing of the relevant constituent. Through $\zeta ^{\alpha }$ we
deform the space, push-forward the fields defined on it, and denote with the
subscript \# the images after $\zeta ^{\alpha }$. We denote with $\mathbf{w}%
_{\alpha }$ the derivative $\dot{\zeta}_{0}^{\alpha }$. If we write $%
\mathfrak{E}\left( \mathfrak{b}\right) =0$ for the balance (\ref{38}),
covariance is the requirement that%
\begin{equation}
\mathfrak{E}_{\#}\left( \zeta _{t}^{\alpha }\left( \mathfrak{b}\right)
\right) \left\vert _{t=0}\right. -\mathfrak{E}\left( \mathfrak{b}\right) =0,
\label{43}
\end{equation}%
for any $\zeta ^{\alpha }$. For the sake of simplicity we assume that the
two observers have the same time scale.

In principle, we presume that each constituent has his peculiar metric $%
\mathbf{g}_{\alpha }$.

\begin{proposition}
Let the explicit representation of $E_{\mathfrak{b}}^{\alpha }$ be given by (%
\ref{38bis}).\ (a) The balance equations (\ref{16}) and (\ref{17}) are
covariant. (b) The symmetric part of $\mathbf{T}_{\alpha }$ in $\mathfrak{b}$
satisfies the Doyle-Ericksen formula%
\begin{equation}
sym\mathbf{T}_{\alpha }=2\rho _{\alpha }\frac{\partial \varepsilon _{\alpha }%
}{\partial \mathbf{g}_{\alpha }}.  \label{43bis}
\end{equation}
\end{proposition}

\subsection{Proof of Proposition 1}

For the proof we follow the program discussed in [10] for the covariance of
balance equations of simple bodies. Since $\mathbf{x}_{\#\alpha }=\zeta
^{\alpha }\left( \mathbf{x,}t\right) $, its time derivative evaluated at $%
t=0 $ gives rise to the following representation for the peculiar velocity
of the $\alpha $th constituent after $\zeta ^{\alpha }$:%
\begin{equation}
\mathbf{x}_{\#\alpha }^{\prime }=\mathbf{x}_{\alpha }^{\prime }+\mathbf{w}%
_{\alpha }\mathbf{.}  \label{44}
\end{equation}%
Under the action of $\zeta ^{\alpha }$, the mass density $\rho _{\alpha }$
and the heat source $r_{\alpha }$ defined on $\mathcal{B}$ remain unchanged,
while vector fields are pushed forward by the tangent map $T\zeta
_{t}^{\alpha }$ which is the identity at $t=0$. A different case occurs for
the internal energy which changes "tensorially" (see [10]). Under the
suggestions of [10], we define%
\begin{equation}
\varepsilon _{\#\alpha }\left( \mathbf{x}_{\#},t\right) =\varepsilon
_{\alpha }\left( \mathbf{x},t,\zeta _{t}^{\alpha \ast }\circ \mathbf{g}%
_{\alpha }\right) ,  \label{45}
\end{equation}%
where $\zeta _{t}^{\alpha \ast }$ denotes pull-back. The meaning of (\ref{45}%
) is discussed in [10] together with the need to parametrize $\varepsilon $
with the metric. Here, we specialize that point of view to the $\alpha $th
constituent and presume that in principle each constituent has its peculiar
metric. In principle $\varepsilon _{\alpha }$\ depends on the $\mathbf{g}%
_{\beta }$'s, $\beta \neq \alpha $, characterizing\ the other constituents.
However, the $\mathbf{g}_{\beta }$'s live on slicings different from the $%
\alpha $th one, then they are not pushed forward by $\zeta ^{\alpha }$\ and
in (\ref{48}) play the r\^{o}le\ of parameters. By chain rule, from (\ref{45}%
) we have%
\begin{equation}
\dot{\varepsilon}_{\#\alpha }=\dot{\varepsilon}_{\alpha }+\frac{\partial
\varepsilon _{\alpha }}{\partial \mathbf{g}_{\alpha }}\cdot L_{\mathbf{w}%
_{\alpha }}\mathbf{g}_{\alpha },  \label{46}
\end{equation}%
where $L_{\mathbf{w}_{\alpha }}$ is the hautonomous Lie derivative of $%
\mathbf{g}_{\alpha }$ following $\mathbf{w}_{\alpha }$. First we write%
\begin{equation*}
\mathfrak{E}_{\#}\left( \zeta _{t}^{\alpha }\left( \mathfrak{b}\right)
\right) =\frac{d}{dt}\int_{\zeta _{t}\left( \mathfrak{b}\right) }\rho
_{\#\alpha }\varepsilon _{\#\alpha }-\int_{\zeta _{t}\left( \mathfrak{b}%
\right) }\rho _{\#}\breve{\varepsilon}_{\#\alpha }-\int_{\zeta _{t}\left( 
\mathfrak{b}\right) }\rho _{\#\alpha }\mathbf{b}_{\#\alpha }\cdot \mathbf{x}%
_{\#\alpha }^{\prime }-
\end{equation*}%
\begin{equation*}
-\int_{\partial \zeta _{t}\left( \mathfrak{b}\right) }\mathbf{T}_{\#\alpha }%
\mathbf{n}_{\#}\cdot \mathbf{x}_{\#\alpha }^{\prime }-\int_{\zeta _{t}\left( 
\mathfrak{b}\right) }\rho _{\#}\mathbf{\check{m}}_{\#\alpha }\cdot \mathbf{x}%
_{\#\alpha }^{\prime }-\int_{\zeta _{t}\left( \mathfrak{b}\right) }\rho _{\#}%
\mathbf{\check{\mu}}_{\#\alpha }\cdot curl_{\#}\mathbf{x}_{\#\alpha
}^{\prime }-
\end{equation*}%
\begin{equation}
-\int_{\partial \zeta _{t}\left( \mathfrak{b}\right) }\mathbf{q}_{\#\alpha
}\cdot \mathbf{n}_{\#}-\int_{\zeta _{t}\left( \mathfrak{b}\right) }\rho
_{\#\alpha }r_{\#\alpha }.  \label{47}
\end{equation}%
By using Gauss theorem and taking into account that $2symgrad\mathbf{x}%
_{\alpha }^{\prime }=L_{\mathbf{w}_{\alpha }}\mathbf{g}_{\alpha }$, and that 
$\breve{\varepsilon}_{\#\alpha }=\breve{\varepsilon}_{\alpha }$ (the
analogous of (\ref{47})), (\ref{43}) becomes%
\begin{equation*}
\mathfrak{E}_{\#}\left( \zeta _{t}^{\alpha }\left( \mathfrak{b}\right)
\right) \left\vert _{t=0}\right. -\mathfrak{E}\left( \mathfrak{b}\right)
=\int_{\mathfrak{b}}\left( \rho _{\alpha }\frac{\partial \varepsilon
_{\alpha }}{\partial \mathbf{g}_{\alpha }}-\frac{1}{2}sym\mathbf{T}_{\alpha
}\right) \cdot L_{\mathbf{w}_{\alpha }}\mathbf{g}_{\alpha }-
\end{equation*}%
\begin{equation}
-\int_{\mathfrak{b}}\left( \rho _{\alpha }\mathbf{b}_{\alpha }+div\mathbf{T}%
_{\alpha }+\rho \mathbf{\check{m}}_{\alpha }\right) \cdot \mathbf{w-}\int_{%
\mathfrak{b}}\left( skw\mathbf{T}_{\alpha }+\rho \mathsf{e}\mathbf{\check{\mu%
}}_{\alpha }\right) \cdot skwgrad\mathbf{w}_{\alpha }=0,  \label{48}
\end{equation}%
where \textsf{e} is the Ricci's alternating symbol and $skw$ extracts the
skew-symmetric part of its argument. The arbitraryness of $\mathbf{w}%
_{\alpha }$ implies that the terms in parentheses must vanish. We then
obtain the balance equations from the last two terms (the last but one
coincides with (\ref{25})). In addition, we have also (\ref{43bis}).$\square 
$

\textbf{Remark}. When there is no exchange of couples between constituents $%
\left( \mathbf{\check{\mu}}_{\alpha }=\mathbf{0}\right) $, and $\varepsilon
_{\alpha }$ doen not depend on the $\mathbf{g}_{\beta }$'s, the Cauchy
stress $\mathbf{T}_{\alpha }$ on $\mathfrak{b}$ measures contact
interactions exerted only by the rest of the $\alpha $th constituent. This
result is has a link with the interpretation given by Williams of mixture
theory in [11, 12], in particular with his "classical interpretation" when
the mixture is not diffusive. In this case, the constituent stress is a
measure of contact interactions between elements of the algebra of parts of
the costituent itself, as in the case of simple bodies: an interpretation
close to principle (ii).

\ \ \ \ 

\textbf{Acknowledgement}. I owe gratitude to Gianfranco Capriz for profound
discussions.

\section{References}

\begin{enumerate}
\item Truesdell, C. A., Rational thermodynamics, Springer Verlag, Berlin,
1984.

\item Noll, W. (1973), Lectures on the foundations of continuum mechanics
and thermodynamics, \emph{Arch. Rational Mech. Anal.}, \textbf{52}, 62-92.

\item Rajagopal, K. R. and Tao, L., \emph{Mechanics of mixtures}, World
Scientific, Singapore, 1995.

\item Tao, L. and Rajagopal, K. R. (1994), On boundary conditions in mixture
theory, in \emph{Recent Advances in Elasticity and Viscoelasticity}, K. R.
Rajagopal ed., World Scientific Publishing, Singapore.

\item Rajagopal, K. R., Wineman, A. S. and Gandhi, M. V. (1986), On boundary
conditions for a certain class of problems in mixture theory, \emph{Int. J.
Engng Sci.}, 29, 1453-1463.

\item Craine, R. E., Green, A. E., Naghdi, P. M. (1970), A mixture of
viscous elastic materials with different constituent temperatures, \emph{%
Quarterly J. Mech. Appl. Math.}, \textbf{23}, 171-184.

\item Capriz, G. (1990), Alcune osservazioni sulla termomeccanica delle
miscele binarie, \emph{Riv. Mat. Univ. Parma}, \textbf{16}, 63-71.

\item M\"{u}ller, I. (1967), A thermodynamic theory of mixtures of fluids, 
\emph{Arch. Rational Mech. Anal.}, \textbf{28}, 1-39.

\item M\"{u}ller, I., \emph{Thermodynamics}, Pitman, 1985.

\item Marsden, J. E. and Hughes, T. J. R., \emph{Mathematical foundations of
elasticity}, Prentice-Hall inc., Englewood Cliffs, 1983.

\item Williams, W. O. (1979), On stresses in mixtures, \emph{Arch. Rational
Mech. Anal.}, \textbf{70}, 251-260.

\item Williams, W. O. (1973), On the theory of mixtures, \emph{Arch.
Rational Mech. Anal.}, \textbf{51}, 239-260.
\end{enumerate}

\end{document}